# Scalar wave scattering by two-layer radial inhomogeneities

Umaporn Nuntaplook[1] and John A. Adam[2]
[1]Department of Mathematics, Mahidol University, Thailand
[2]Department of Mathematics & Statistics, Old Dominion University, Norfolk, Virginia, USA

## 1. Introduction

It is known that the Jost-function formulation of quantum scattering theory can be applied to classical problems concerned with the scattering of a plane scalar wave by a medium with a spherically symmetric inhomogeneity of finite extent. We have applied this technique to solve the radial differential equation for the scattering from a constant spherical inhomogeneity and a piecewise constant by two-layer spherical inhomogeneity. This could represent a spherical scatterer with a piecewise increasing or decreasing refractive index, for example. When the problem cannot be solved analytically in closed form, the Jost integral formula can be used to convert it into an integral equation with the corresponding boundary conditions. An iteration procedure can be used to solve the $l = 0$ Jost integral equation for piecewise increasing or decreasing refractive indices.

## 2. Governing equations

The equation of interest is the radial equation for potential $V(r)$, where in the following notation, the energy $E = \sqrt{k}$ and $l$ is the angular momentum parameter:

$$\frac{d^2 u_l(r)}{dr^2} + [k^2 - \frac{l(l+1)}{r^2} - V(r)]u_l(r) = 0. \qquad 1$$

For the following analysis to hold, we must impose certain general requirements on $V(r)$: it must be a real function vanishing at $r = \infty$, and must be almost everywhere continuous; furthermore, we require that [1]

$$\int_c^\infty |V(r)| dr = M(c) < \infty;$$

$$\int_0^{c'} r|V(r)| dr = N(c') < \infty,$$

where $c$ and $c'$ are arbitrary constants greater than zero. This conditions are readily satisfied for non-singular $V(r)$ with compact support. The relevant boundary conditions are discussed below, based on the development reported in [2] (see also [3]).

**Jost boundary conditions at** $r = 0$:

For small $r$,

$$|-l(l+1)/r^2| >> |k^2 - V(r)|,$$

so by neglecting the term $[k^2 - V(r)]$. equation (1) becomes

$$\frac{d^2 u_l(r)}{dr^2} - [\frac{l(l+1)}{r^2}]u_l(r) = 0, \qquad 2$$

which has a regular singular point at $r = 0$ with the general solution

$$u_l(r) = \alpha r^{l+1} + \beta r^{-l}. \qquad 3$$

We define two linearly independent solutions of (2) with the behavior

$$\phi(r) = r^{l+1}[1 + o(1)],$$

$$\phi_1(r) = r^{-l}[1 + o(1)],$$

$o$ being the order symbol. We also define $\lambda = l + \frac{1}{2}$, so that (1) becomes even in $\lambda$:

$$\frac{d^2u_{\lambda-\frac{1}{2}}(r)}{dr^2} + [k^2 - \frac{\lambda^2 - \frac{1}{4}}{r^2} - V(r)]u_{\lambda-\frac{1}{2}}(r) = 0. \quad 4$$

We now consider the solutions of (4) as a function of the parameters $\lambda$ and $k$, which in the general case may be complex variables. Thus $\phi(r) \to \phi_1(r)$ as $\lambda \to -\lambda$, and we therefore replace $\phi(r)$ and $\phi_1(r)$ by $\phi(\lambda, k, r)$ and $\phi(-\lambda, k, r)$, respectively. We write $u_{\lambda-\frac{1}{2}}(r)$ instead of $u_l(r)$ in order that the notation be consistent. We solve (4) exactly by converting it into an integral equation using the method of variation of parameters with the boundary conditions. We thus obtain

$$\phi(\lambda, k, r) = r^{\lambda+\frac{1}{2}} + \frac{1}{2}\lambda^{-1}\int_0^r [(\xi/r)^\lambda - (r/\xi)^\lambda] \times (r\xi)^{\frac{1}{2}}[k^2 - V(\xi)]\phi(\lambda, k, \xi)d\xi. \quad 5$$

From equation (5) we can write the Jost boundary conditions at $r = 0$ as

$$\lim_{r\to 0}\phi(\lambda, k, r) = 0; \quad 6a$$

$$\lim_{r\to 0}\frac{d\phi(\lambda, k, r)}{dr} = \lim_{r\to 0}(\lambda + \frac{1}{2})r^{\lambda-\frac{1}{2}}. \quad 6b$$

Equation (5) is a Volterra integral equation, and we write its solution as a perturbation expansion

$$\phi(\lambda, k, r) = \sum_{n=0}^{\infty}\phi_n(\lambda, k, r), \quad 7$$

where

$$\phi_0(\lambda, k, r) = r^{\lambda+\frac{1}{2}},$$

and

$$\phi_{n+1}(\lambda, k, r) = \frac{1}{2}\lambda^{-1}\int_0^r [(\xi/r)^\lambda - (r/\xi)^\lambda](r\xi)^{\frac{1}{2}} \times [k^2 - V(\xi)]\phi_n(\lambda, k, \xi)d\xi. \quad 8$$

The perturbation expansion for $\phi(\lambda, k, r)$is bounded term by term and is unrestrictedly convergent. Since the Wronskian of the two linearly independent solutions of (4) is nonzero and constant, we can evaluate the Wronskian of $\phi(\lambda, k, r)$ and $\phi(-\lambda, k, r)$ with $r^{\lambda+\frac{1}{2}}$:

$$W[\phi(\lambda, k, r), \phi(-\lambda, k, r)] = \phi(\lambda, k, r)\phi'(-\lambda, k, r) - \phi'(\lambda, k, r)\phi(-\lambda, k, r) = -2\lambda. \quad 9$$

**Jost boundary conditions as $r \to \infty$**

For large $r$, we can neglect the term

$$[-\frac{\lambda^2 - \frac{1}{4}}{r^2} - V(r)],$$

in (4), which therefore becomes

$$\frac{d^2u_{\lambda-\frac{1}{2}}(r)}{dr^2} + k^2u_{\lambda-\frac{1}{2}}(r) = 0. \quad 10$$

Equation (10) has the general solution

$$u_{\lambda-\frac{1}{2}}(r) = \alpha e^{-ikr} + \beta e^{ikr}. \quad 11$$

We can construct a solution $f(\lambda, k, r)$ (the Jost solution) with the asymptotic behavior

$$\lim_{r\to\infty} e^{ikr} f(\lambda,k,r) = 1. \quad 12$$

We use the method of variation of parameters so that the integral equation for $f(\lambda,k,r)$ with the asymptotic condition (12) becomes

$$f(\lambda,k,r) = e^{-ikr} + k^{-1}\int_r^{\infty}[\sin k(r'-r)] \times [V(r') + (\lambda^2 - \frac{1}{4})(r')^{-2}]f(\lambda,k,r')dr'. \quad 13$$

We write the solution of (13) as a perturbation expansion

$$f(\lambda,k,r) = \sum_{n=0}^{\infty} g_n(\lambda,k,r), \quad 14$$

where

$$g_0 = e^{-ikr},$$

and

$$g_{n+1}(\lambda,k,r) = k^{-1}\int_r^{\infty}[\sin k(r'-r)] \times [V(r'^2 - \frac{1}{4})(r'^{-2}]g_n(\lambda,k,r')dr'. \quad 15$$

The perturbation expansion for $f(\lambda,k,r)$ is bounded for any $\lambda$. We state the following relation:

$$W[f(\lambda,k,r), f(\lambda,-k,r)] = 2ik, \quad 16$$

where we have evaluated the Wronskian by substituting for $f(\lambda,\pm k,r)$ its asymptotic behavior, $e^{\pm ikr}$.

## 3. Scattering From A Piecewise Constant by Multi-Layer Spherically Symmetric Inhomogeneities

We are now in a position to apply the method outlined above to the problem of scattering from a piecewise constant in a multi-layer spherical inhomogeneities. For a three-layer inhomogeneity we define the following potential

$$\begin{aligned} &Region\ 1: V(r) = -V_1, k(r) = k_1, r < R_1;\\ &Region\ 2: V(r) = -V_2, k(r) = k_2, R_1 < r < R_2;\\ &Region\ 3: V(r) = 0, k(r) = k, r < R_2. \end{aligned} \quad 17$$

The solutions in the three regions are:

$$\begin{aligned} &Region\ 1: u^{(1)}_{\lambda-\frac{1}{2}}(k,r) = r[Aj_{\lambda-\frac{1}{2}}(k_1 r) + By_{\lambda-\frac{1}{2}}(k_1 r)];\\ &Region\ 2: u^{(2)}_{\lambda-\frac{1}{2}}(k,r) = r[Cj_{\lambda-\frac{1}{2}}(k_2 r) + Dy_{\lambda-\frac{1}{2}}(k_2 r)];\\ &Region\ 3: u^{(3)}_{\lambda-\frac{1}{2}}(k,r) = r[Eh^{(1)}_{\lambda-\frac{1}{2}}(kr) + Fh^{(2)}_{\lambda-\frac{1}{2}}(kr)]. \end{aligned} \quad 18$$

where again $j_{\lambda-\frac{1}{2}}(k_1 r)$, $y_{\lambda-\frac{1}{2}}(k_2 r)$, $h^{(1)}_{\lambda-\frac{1}{2}}(kr)$, and $h^{(2)}_{\lambda-\frac{1}{2}}(kr)$ are spherical Bessel, Neumann, and Hankel functions of the first kind and second kind, respectively. Choosing $u^{(1)}_{\lambda-\frac{1}{2}}(k_1 r)$ to be $\phi(\lambda,k,r)$ and imposing the boundary conditions (6a,6b) at $r = 0$, we find that $B = 0$ and

$$\begin{aligned} &\phi_1(\lambda,k,r) = 2^{\lambda+\frac{1}{2}}\pi^{-\frac{1}{2}}k_1^{-\lambda+\frac{1}{2}}\Gamma(\lambda+1)rj_{\lambda-\frac{1}{2}}(k_1 r),\\ &\phi_1'(\lambda,k,r) = 2^{\lambda+\frac{1}{2}}\pi^{-\frac{1}{2}}k_1^{-\lambda+\frac{1}{2}}\Gamma(\lambda+1) \times [j_{\lambda-\frac{1}{2}}(k_1 r) + k_1 r j'_{\lambda-\frac{1}{2}}(k_1 r)], \end{aligned} \quad 19$$

where the prime denotes differentiation with respect to the argument of the function, $\Gamma$ is the gamma function, and we have used the following series representation for $j_{\lambda-\frac{1}{2}}(k_1 r)$ [Handbook of Mathematical Functions (McGraw Hill Book, p. 263)]:

$$j_{\lambda-\frac{1}{2}}(k_1 r) = \sum_{n=0}^{\infty} \frac{(-1)^n \pi^{\frac{1}{2}} (k_1 r/2)^{\lambda+2n-\frac{1}{2}}}{2n!\Gamma(\lambda+n+1)}, \ \lambda - \frac{1}{2} \neq -1, -2, -3, \ldots \qquad 20$$

Choosing $u^{(2)}_{\lambda-\frac{1}{2}}(k_2 r)$ to be $\phi_2(\lambda, k, r)$ and imposing the continuity at the boundary $r = R_1$ by matching the continuity of $\phi_1$ with $\phi_2$ and $\phi_1'$ with $\phi_2'$, we have

$$\phi_2(\lambda, k, r) = r[C j_{\lambda-\frac{1}{2}}(k_2 r) + D y_{\lambda-\frac{1}{2}}(k_2 r)],$$
$$\phi_2'(\lambda, k, r) = C[j_{\lambda-\frac{1}{2}}(k_2 r) + k_2 r j'_{\lambda-\frac{1}{2}}(k_2 r)] + D[y_{\lambda-\frac{1}{2}}(k_2 r) + k_2 r y'_{\lambda-\frac{1}{2}}(k_2 r)], \qquad 21$$

where

$$C = -\frac{m(j_{\lambda-\frac{1}{2}}(k_1 R_1) y'_{\lambda-\frac{1}{2}}(k_2 R_1) - \frac{k_1}{k_2} j'_{\lambda-\frac{1}{2}}(k_1 R_1) y_{\lambda-\frac{1}{2}}(k_2 R_1))}{j'_{\lambda-\frac{1}{2}}(k_2 R_1) y_{\lambda-\frac{1}{2}}(k_2 R_1) - j_{\lambda-\frac{1}{2}}(k_2 R_1) y'_{\lambda-\frac{1}{2}}(k_2 R_1)};$$
$$D = \frac{m(j_{\lambda-\frac{1}{2}}(k_1 R_1) j'_{\lambda-\frac{1}{2}}(k_2 R_1) - \frac{k_1}{k_2} j'_{\lambda-\frac{1}{2}}(k_1 R_1) j_{\lambda-\frac{1}{2}}(k_2 R_1))}{j'_{\lambda-\frac{1}{2}}(k_2 R_1) y_{\lambda-\frac{1}{2}}(k_2 R_1) - j_{\lambda-\frac{1}{2}}(k_2 R_1) y'_{\lambda-\frac{1}{2}}(k_2 R_1)};$$

$$\text{and } m = 2^{\lambda+\frac{1}{2}} \pi^{-\frac{1}{2}} k_1^{-\lambda+\frac{1}{2}} \Gamma(\lambda+1). \qquad 22$$

Choosing $u^{(3)}_{\lambda-\frac{1}{2}}(kr)$ to be $f(\lambda, k, r)$ and imposing the Jost solution condition (12) at infinity we find that $E = 0$, $F = k e^{-i\frac{\pi}{2}(\lambda+\frac{1}{2})}$, and hence

$$f(\lambda, k, r) = k e^{-i\frac{\pi}{2}(\lambda+\frac{1}{2})} r h^{(2)}_{\lambda-\frac{1}{2}}(kr),$$
$$f'(\lambda, k, r) = k e^{-i\frac{\pi}{2}(\lambda+\frac{1}{2})} [h^{(2)}_{\lambda-\frac{1}{2}}(kr) + kr h^{(2)\prime}_{\lambda-\frac{1}{2}}(kr)], \qquad 23$$

where we have used the following asymptotic form for $h^{(2)}_{\lambda-\frac{1}{2}}(kr)$:

$$\lim_{kr\to\infty} h^{(2)}_{\lambda-\frac{1}{2}}(kr) = \frac{1}{kr} e^{-i[kr - \frac{\pi}{2}(\lambda+\frac{1}{2})]}.$$

Since the point $r = R_2$ is the common domain of $\phi_2(\lambda, k, r)$ and $f(\lambda, k, r)$, we evaluate the Jost function at $r = R_2$ and thus obtain

$$\begin{aligned} f(\lambda, k) &= W[f(\lambda, k, r), \phi_2(\lambda, k, r)]_{r=R_2} \\ &= f(\lambda, k, r)\phi_2'(\lambda, k, r) - f'(\lambda, k, r)\phi_2(\lambda, k, r) \\ &= 2^{\lambda+\frac{1}{2}} \pi^{-\frac{1}{2}} \Gamma(\lambda+1) k_1^{-\lambda+\frac{1}{2}} k e^{-i\frac{\pi}{2}(\lambda+\frac{1}{2})} R_2^2 \\ &\times \{h^{(2)}_{\lambda-\frac{1}{2}}(kR_2) k_2 [a_1 j_{\lambda-\frac{1}{2}}(k_1 R_1) + \frac{k_1}{k_2} a_3 j'_{\lambda-\frac{1}{2}}(k_1 R_1)] \\ &- h^{(2)\prime}_{\lambda-\frac{1}{2}}(kR_2) k [a_2 j_{\lambda-\frac{1}{2}}(k_1 R_1) + \frac{k_1}{k_2} a_4 j'_{\lambda-\frac{1}{2}}(k_1 R_1)]\} \\ &/[j'_{\lambda-\frac{1}{2}}(k_2 R_1) y_{\lambda-\frac{1}{2}}(k_2 R_1) - j_{\lambda-\frac{1}{2}}(k_2 R_1) y'_{\lambda-\frac{1}{2}}(k_2 R_1)]. \end{aligned} \qquad 24$$

We also have that

$$
\begin{aligned}
f(\lambda,-k) &= 2^{\lambda+\frac{1}{2}}\pi^{-\frac{1}{2}}\Gamma(\lambda+1)k_1^{-\lambda+\frac{1}{2}}ke^{-i\frac{\pi}{2}(\lambda+\frac{1}{2})}R_2^2e^{i\pi(\lambda-\frac{1}{2})} \\
&\times\{-h^{(1)}_{\lambda-\frac{1}{2}}(kR_2)k_2[a_1j_{\lambda-\frac{1}{2}}(k_1R_1)+\frac{k_1}{k_2}a_3j'_{\lambda-\frac{1}{2}}(k_1R_1)] \\
&+h^{(1)'}_{\lambda-\frac{1}{2}}(kR_2)k[a_2j_{\lambda-\frac{1}{2}}(k_1R_1)+\frac{k_1}{k_2}a_4j'_{\lambda-\frac{1}{2}}(k_1R_1)]\} \\
&/[j'_{\lambda-\frac{1}{2}}(k_2R_1)y_{\lambda-\frac{1}{2}}(k_2R_1)-j_{\lambda-\frac{1}{2}}(k_2R_1)y'_{\lambda-\frac{1}{2}}(k_2R_1)], \qquad 25
\end{aligned}
$$

where we have used the following identities:

$$
h^{(2)}_{\lambda-\frac{1}{2}}(kre^{i\pi})=h^{(2)}_{\lambda-\frac{1}{2}}(-kr)=(-1)^{\lambda-\frac{1}{2}}h^{(1)}_{\lambda-\frac{1}{2}}(kr)=e^{i\pi(\lambda-\frac{1}{2})}h^{(1)}_{\lambda-\frac{1}{2}}(kr);
$$

$$
h^{(2)'}_{\lambda-\frac{1}{2}}(-kr)=(-1)^{\lambda+\frac{1}{2}}h^{(1)'}_{\lambda-\frac{1}{2}}(kr)=e^{i\pi(\lambda+\frac{1}{2})}h^{(1)'}_{\lambda-\frac{1}{2}}(kr)=-e^{i\pi(\lambda-\frac{1}{2})}h^{(1)'}_{\lambda-\frac{1}{2}}(kr),\ \lambda-\frac{1}{2}=0,1,2,\ldots.
$$

The $S$-matrix is then given by

$$
\begin{aligned}
S(\lambda,k) &= -\{kh^{(2)'}_{\lambda-\frac{1}{2}}(kR_2)[a_2j_{\lambda-\frac{1}{2}}(k_1R_1)+\frac{k_1}{k_2}a_4j'_{\lambda-\frac{1}{2}}(k_1R_1)] \\
&-k_2h^{(2)}_{\lambda-\frac{1}{2}}(kR_2)[a_1j_{\lambda-\frac{1}{2}}(k_1R_1)+\frac{k_1}{k_2}a_3j'_{\lambda-\frac{1}{2}}(k_1R_1)]\} \\
&/\{kh^{(1)'}_{\lambda-\frac{1}{2}}(kR_2)[a_2j_{\lambda-\frac{1}{2}}(k_1R_1)+\frac{k_1}{k_2}a_4j'_{\lambda-\frac{1}{2}}(k_1R_1)] \\
&-k_2h^{(1)}_{\lambda-\frac{1}{2}}(kR_2)[a_1j_{\lambda-\frac{1}{2}}(k_1R_1)+\frac{k_1}{k_2}a_3j'_{\lambda-\frac{1}{2}}(k_1R_1)]\}, \qquad 26
\end{aligned}
$$

where

$$
\begin{aligned}
a_1 &= j'_{\lambda-\frac{1}{2}}(k_2R_1)y'_{\lambda-\frac{1}{2}}(k_2R_2)-y'_{\lambda-\frac{1}{2}}(k_2R_1)j'_{\lambda-\frac{1}{2}}(k_2R_2); \\
a_2 &= j'_{\lambda-\frac{1}{2}}(k_2R_1)y_{\lambda-\frac{1}{2}}(k_2R_2)-j_{\lambda-\frac{1}{2}}(k_2R_2)y'_{\lambda-\frac{1}{2}}(k_2R_1); \\
a_3 &= y_{\lambda-\frac{1}{2}}(k_2R_1)j'_{\lambda-\frac{1}{2}}(k_2R_2)-j_{\lambda-\frac{1}{2}}(k_2R_1)y'_{\lambda-\frac{1}{2}}(k_2R_2); \\
a_4 &= j_{\lambda-\frac{1}{2}}(k_2R_1)y_{\lambda-\frac{1}{2}}(k_2R_2)-y_{\lambda-\frac{1}{2}}(k_2R_1)j_{\lambda-\frac{1}{2}}(k_2R_2).
\end{aligned}
$$

We can calculate the Jost function for $\lambda=\frac{1}{2}$ from (24):

$$
\begin{aligned}
f(\frac{1}{2},k) &= \frac{1}{4}e^{-ikR_2}\{([\frac{(k-ik_2)}{k_1}+\frac{(ik+k_2)}{k_2}]e^{k_2(R_2-R_1)}+[\frac{(k+ik_2)}{k_1}+\frac{(k_2-ik)}{k_2}]e^{-k_2(R_2-R_1)})e^{ik_1R_1} \\
&+([\frac{(ik+k_2)}{k_2}-\frac{(k-ik_2)}{k_1}]e^{k_2(R_2-R_1)}+[\frac{(k_2-ik)}{k_2}-\frac{(k+ik_2)}{k_1}]e^{-k_2(R_2-R_1)})e^{-ik_1R_1}\}. \qquad 27
\end{aligned}
$$

where we have used the following relations:

$$
j_0(kR)=\sin kR/(kR),\ j'_0(kR)=\cos kR/(kR)-[\sin kR/(kR)^2],
$$

$$
h^{(2)}_0(kR)=-e^{-ikR}/(ikR),\ h^{(2)'}_0(kR)=e^{-ikR}[1+1/(ikR)]/(kR).
$$

## 4. The Jost integral equation for $\lambda=\frac{1}{2}$ and some approximate solutions for the three-layer model

We now apply the perturbation expansion method in Section 2 to the case of scattering from a piecewise constant by multi-layer spherical inhomogeneity. We have already calculated $f(\frac{1}{2},k)$ exactly in equation (27). In another article we will use the exact solution of the Jost function to report on the accuracy of the iteration procedure. If we assume there is an $R$ such that $V(r)=0$ for $r>R$ (certainly true in optics!) and let

$$g(\frac{1}{2},k,r) = e^{ikr}f(\frac{1}{2},k,r), \qquad 28$$

then (15) becomes the Jost integral equation for $\lambda = \frac{1}{2}$:

$$g(\frac{1}{2},k,r) = 1 + (2ik)^{-1}\int_r^R [1 - e^{2ik(r-r')}]V(r')g(\frac{1}{2},k,r')dr' \qquad 29a$$

$$= 1 - V_1(2ik)^{-1}\int_r^{R_1}[1 - e^{2ik(r-r')}]g(\frac{1}{2},k,r')dr', \qquad 29b$$

for the potential defined by region I in equation (17). Next we write the solution of (29) as a perturbation expansion

$$g(\frac{1}{2},k,r) = \sum_{n=0}^{\infty} g_n(\frac{1}{2},k,r), \qquad 30$$

where

$$g_0(\frac{1}{2},k,r) = 1$$

and

$$g_n(\frac{1}{2},k,r) = 1 + (2ik)^{-1}\int_r^R[1 - e^{2ik(r-r')}]V(r')g_{n-1}(\frac{1}{2},k,r')dr'. \qquad 31$$

From (6a/6b), we have

$$\lim_{r\to 0}\phi(\frac{1}{2},k,r) = 0, \qquad 32a$$

$$\lim_{r\to 0}\frac{\phi'(\frac{1}{2},k,r)}{dr} = 1. \qquad 32b$$

Now $f(\frac{1}{2},k,r)$ and $f'(\frac{1}{2},k,r)$ are finite and we can evaluate $f(\frac{1}{2},k)$ at $r = 0$ using (32a/b), thus obtaining the useful relation

$$f(\frac{1}{2},k) = f(\frac{1}{2},k,0) = g(\frac{1}{2},k,0). \qquad 33$$

The first iteration $g_I(\frac{1}{2},k,0)$ of (29b) is

$$g_I(\frac{1}{2},k,0) = g_0(\frac{1}{2},k,0) + g_1(\frac{1}{2},k,0)$$

$$= 1 - \frac{1}{4}\{[(\frac{k_1}{k})^2 - 1][1 - \cos 2kR_1] + [1 - (\frac{k_2}{k})^2][\cos 2kR_2 - \cos 2kR_1]\}$$

$$+ \frac{i}{2}\{[(\frac{k_1}{k})^2 - 1][kR_1 - \frac{1}{2}\sin 2kR_1] + [1 - (\frac{k_2}{k})^2][k(R_2 - R_1) - \frac{1}{2}(\sin 2kR_2 - \sin 2kR_1]\}. \qquad 34$$

The second iteration $g_{II}(\frac{1}{2},k,0)$ is

$$
\begin{aligned}
g_{II}(\tfrac{1}{2},k,0) &= g_0(\tfrac{1}{2},k,0) + g_1(\tfrac{1}{2},k,0) ++ g_2(\tfrac{1}{2},k,0) \\
&= 1 + \frac{1}{4}\{[(\frac{k_1}{k})^2 - 1][\cos 2kR_1 - 1] + [1 - (\frac{k_2}{k})^2][\cos 2kR_2 - \cos 2kR_1]\} \\
&- \frac{1}{8}\{[(\frac{k_1}{k})^2 - 1]^2(kR_1[kR_1 + \sin 2kR_1] + \frac{3}{2}[\cos 2kR_1 - 1]) \\
&+ [(\frac{k_1}{k})^2 - 1][1 - (\frac{k_2}{k})^2](k(R_2 - R_1)[2kR_1 - k(R_2 - R_1) - \sin 2kR_1] \\
&- [\cos 2k(R_2 - R_1) - 1] + \frac{3}{2}[\cos 2kR_2 - \cos 2kR_1] \\
&+ k(R_2 - R_1)[\sin 2kR_2 + \sin 2kR_1] + kR_1[\sin 2kR_2 - \sin 2kR_1]) \\
&+ [1 - (\frac{k_2}{k})^2]^2(k(R_2 - R_1)[2k(R_2 - R_1) - \sin 2k(R_2 - R_1) \\
&+ \sin 2kR_2 - \sin 2kR_1] + [\cos 2k(R_2 - R_1) - 1])\} \\
&+ i\{\frac{1}{2}([(\frac{k_1}{k})^2 - 1][kR_1 - \frac{1}{2}\sin 2kR_1] \\
&+ [1 - (\frac{k_2}{k})^2][k(R_2 - R_1) - \frac{1}{2}[\sin 2kR_2 - \sin 2kR_1]]) \\
&- \frac{1}{8}([(\frac{k_1}{k})^2 - 1]^2(kR_1[\cos 2kR_1 + 2] - \frac{3}{2}\sin 2kR_1) \\
&+ [(\frac{k_1}{k})^2 - 1][1 - (\frac{k_2}{k})^2](kR_1[\cos 2kR_2 - \cos 2kR_1] \\
&+ k(R_2 - R_1)[\cos 2kR_2 + \cos 2kR_1] \\
&- k(R_2 - R_1)[\cos 2kR_1 - 1] - \frac{3}{2}[\sin 2kR_2 - \sin 2kR_1]) \\
&- [1 - (\frac{k_2}{k})^2]^2(k(R_2 - R_1)[[\cos 2k(R_2 - R_1) - \cos 2kR_2 + \cos 2kR_1]]))\}. \qquad 35
\end{aligned}
$$

For real $\lambda$ and $k$, we have

$$f(\lambda, -k) = f^*(\lambda, k), \qquad 36$$

and therefore

$$
\begin{aligned}
\sigma_0/\pi R_1^2 &= |1 - e^{2i\delta(\frac{1}{2},k)}|^2/(kR_1)^2 \\
&= |1 - [f(\tfrac{1}{2},k)/f^*(\tfrac{1}{2},k)]|^2/(kR_1)^2, \qquad 37
\end{aligned}
$$

where $\sigma_0$ is the $l = 0$ total cross section. The accuracy of these approximations as functions of $k, k_1$ and $k_2$ will be reported elsewhere.

## 5. Jost integral equation formulation for arbitrary $\lambda$: 3-layer model

In case of scattering from a piecewise constant spherical inhomogeneity, the two integral equations (5) and (13) become:

$$
\begin{aligned}
\phi(\lambda,k,r) &= r^{\lambda+\frac{1}{2}} + \frac{1}{2}\lambda^{-1}\int_0^{R_1}[(\xi/r)^\lambda - (r/\xi)^\lambda] \times (r\xi)^{\frac{1}{2}}[k^2 + V_1]\phi(\lambda,k,\xi)d\xi \\
&+ \frac{1}{2}\lambda^{-1}\int_{R_1}^{R_2}[(\xi/r)^\lambda - (r/\xi)^\lambda] \times (r\xi)^{\frac{1}{2}}[k^2 + V_2]\phi(\lambda,k,\xi)d\xi \\
&+ \frac{1}{2}\lambda^{-1}\int_{R_2}^{r}[(\xi/r)^\lambda - (r/\xi)^\lambda](r\xi)^{\frac{1}{2}}k^2\phi(\lambda,k,\xi)d\xi; \qquad 38
\end{aligned}
$$

$$
\begin{aligned}
f(\lambda,k,r) = e^{-ikr} &+ k^{-1}\int_{r}^{R_1}[\sin k(r'-r)] \times [-V_1 + (\lambda^2 - \tfrac{1}{4})/(r'^2)]f(\lambda,k,r')dr' \\
&+ k^{-1}\int_{R_1}^{R_2}[\sin k(r'-r)] \times [-V_2 + (\lambda^2 - \tfrac{1}{4})/(r'^2)]f(\lambda,k,r')dr' \\
&+ k^{-1}\int_{R_2}^{\infty}[\sin k(r'^2 - \tfrac{1}{4})/(r'^2)]f(\lambda,k,r')dr'. \qquad (39)
\end{aligned}
$$

## 6. Summary

This iterative technique may be most useful when the scattering system is more complicated than those discussed here. By comparing the present formulation with the numerical results obtained for a constant spherical inhomogeneity [2], it appears that the iteration technique is good for problems with long wavelengths ($kR_1 << 1$) for any $k_1/k$. For shorter wavelengths, small $k_1/k$ (e.g., $k_1/k = 1.1$) gives a good approximation to $\sigma_0$ for the entire range of $kR_1$ considered ($0 \leq R_1 \leq 2\pi$); however, large $k_1/k$ (e.g., $k_1/k = 1.5, 2.0$) gives a good approximation to $\sigma_0$ in the range of $0 < kR_1 < 3\pi/4$. In case of a piecewise constant spherical inhomogeneity, the iteration procedure gives a better approximation for the problem with long wavelengths ($kR_1 << 1$) only for small ratios of $k_1/k$ and $k_2/k$ (e.g., $k_1/k = 0.7, k_2/k = 0.9; k_1/k = 1.1, k_2/k = 1.3$). For a larger $k_1/k$ and $k_2/k$ (e.g., $k_1/k = 1.5, k_2/k = 1.2$), it gives a good approximation when $kR_1 < 2\pi/3$. The approximation for the Jost function becomes less accurate for larger rations of wavenumber $k_1/k$ and $k_2/k$ (e.g., $k_1/k = 2.0, k_2/k = 1.5$). When the ratios of wavenumbers $k_1/k$ is greater than $k_2/k$, we have a better approximation. However, the approximation for the Jost function is still better than the total cross section for the large wavelengths. For shorter wavelengths, all rations of the wavenumbers give a better approximation to $\sigma_0$ for approximately $kR_1 > 2\pi/3$ [4]. These results will be reported in more detail elsewhere.